\documentstyle[aps,prd]{revtex}
\draft
\begin{document}
\thispagestyle{empty}
\thispagestyle{empty}
{\baselineskip0pt
\leftline{\large\baselineskip16pt\sl\vbox to0pt{\hbox{DAMTP} 
               \hbox{University of Cambridge}\vss}}
\rightline{\large\today}
}
\vskip20mm
\begin{center}
{\Large\bf The Einstein Equations on the 3-Brane World}
\end{center}

\begin{center}
{\large Tetsuya Shiromizu${}^{1,3,4}$, Kei-ichi Maeda${}^{2,5}$ 
and Misao Sasaki${}^{2,3,6}$}
\vskip 3mm
\sl{${}^1$DAMTP, University of Cambridge \\ 
Silver Street, Cambridge CB3 9EW, United Kingdom
\vskip 5mm
${}^2$Isaac Newton Institute, University of Cambridge, \\
20 Clarkson Road, Cambridge CB3 0EH, United Kingdom
\vskip 5mm
${}^3$Department of Physics, The University of Tokyo, Tokyo 113-0033, Japan
\vskip 5mm
${}^4$Research Centre for the Early Universe(RESCEU), \\ 
The University of Tokyo, Tokyo 113-0033, Japan
\vskip 5mm 
${}^5$Department of Physics, Waseda University, Shinjuku, 
Tokyo 169-8555, Japan
\vskip 5mm 
${}^6$Department of Earth and Space Science, Graduate School of Science,\\
Osaka University, Toyonaka 560-0043, Japan}
\end{center}


\begin{abstract} 
We carefully investigate the gravitational equations of the brane world,
in which all the matter forces except gravity are confined on the
3-brane in a 5-dimensional spacetime with $Z_2$ symmetry.
 We derive the effective gravitational equations on the brane,
which reduce to the conventional Einstein equations in the low energy
limit. {}From our general argument we conclude that the first 
Randall \& Sundrum-type theory (RS1) [hep-ph/9905221] predicts that 
the brane with the negative tension 
is an anti-gravity world and hence should be 
excluded from the physical point of view. Their second-type theory (RS2)
[hep-th/9906064] where the brane has the positive tension 
provides the correct signature of gravity. 
In this latter case, if the bulk spacetime is exactly anti-de Sitter,
generically the matter on the brane is required to be spatially
homogeneous because of the Bianchi identities.
By allowing deviations from anti-de Sitter in the bulk,
the situation will be relaxed and the Bianchi identities 
give just the relation between the Weyl tensor and the energy momentum
tensor. 
In the present brane world scenario, the effective Einstein equations
cease to be valid during an era when the cosmological constant
on the brane
is not well-defined, such as in the case of the matter dominated by
the potential energy of the scalar field.
\end{abstract}

\pacs{PACS: 04.50.+h; 98.80.Cq\qquad
DAMTP-1999-150;
NI99018-SFU;
UTAP-349;
RESCEU-40/99;\\ 
WU-AP/85/99;OUTAP-103
}

\vskip1cm


Recent progress in the superstring theory tells us that we are living 
in 11 dimensions\cite{String}, and different string theories are
connected with each other via dualities. Among string theories, 
the 10-dimensional $E_8 \times E_8$ heterotic string theory
is a strong candidate for our real world 
because the theory may contain the standard model. Recently Horava \&
Witten showed that the 10-dimensional $E_8 \times E_8$ heterotic string
 is related to an 11-dimensional theory on the orbifold ${\bf R}^{10}
\times {\bf S}^1/Z_2$ \cite{Witten}. 
Therein the standard model particles are confined to the
4-dimensional spacetime. On the other hand, gravitons propagate
in the full spacetime. 

This situation can be simplified to a 5-dimensional problem where matter
fields are confined to the 4-dimensional spacetime while gravity 
acts in 5 dimensions.
 In this category much work has been done.
Among of all, the pioneer work in spacetime with 
one extra dimension has been 
done by Randall \& Sundrum\cite{RS1,RS2}\cite{early} where our brane is 
identical to a domain wall in 5-dimensional anti-de Sitter 
spacetime. In their first paper\cite{RS1}, they 
proposed a mechanism to solve the hierarchy problem by a small 
extra dimension. 
In their second paper\cite{RS2}, the brane world with a positive
tension was investigated. Then a non-perturbative aspect of the theory
was investigated\cite{SUGRA}. The final fate of 
gravitational collapse was discussed in the brane world picture\cite{BH}.  
The inflation solution has been discovered\cite{Kaloper,Nihei,Kim}. 
In these treatments, however, 
the contribution from matter excitations has not been seriously 
considered. Such work is partially performed in a cosmological 
context linked to the conventional Friedman equation
\cite{BDL,Csaba,Cline,Cvetic}.  The cosmological solution
associated with the heterotic string theory also has been
constructed\cite{real}.  We mention 
work on the brane world motivated by the hierarchy problem. 
Before Randall \& Sundrum's work, large 
extra dimensions were proposed to solve the hierarchy problem\cite{MM}. 
The related cosmology also has been actively investigated\cite{cosmos}. 

In this paper, we derive the effective Einstein equations on the
3-brane. For simplicity the bulk spacetime is assumed to have 
5 dimensions. In the beginning we do not assume any conditions on the
bulk spacetime. Later, we assume the $Z_2$-symmetry and confinement of
the matter energy momentum tensor on the brane, in accordance with the
brane world scenario based on the 
Horava \& Witten theory\cite{Witten}. The notation basically follows 
Wald's text\cite{wald}. 

In the brane world scenario, our 4-dimensional world is 
described by a domain wall (3-brane) $(M,q_{\mu\nu})$ in 5-dimensional
spacetime $(V,g_{\mu\nu})$. We denote the vector unit normal to $M$
by $n^\alpha$ and the induced metric on $M$ by 
$q_{\mu\nu} = g_{\mu\nu} - n_{\mu}n_{\nu}$.
We start with the Gauss equation,
\begin{equation}
{}^{(4)}R^\alpha_{~\beta\gamma\delta}
= {}^{(5)}R^\mu_{~\nu\rho\sigma}
q^{~\alpha}_\mu q_\beta^{~\nu} q_\gamma^{~\rho}
 q_\delta^{~\sigma} + K^\alpha_{~\gamma}K_{\beta\delta} 
-K^\alpha_{~\delta}K_{\beta\gamma}\,, 
\label{Gauss}
\end{equation}
and the Codacci equation,
\begin{equation}
D_\nu K^{~\nu}_\mu - D_\mu K 
= {}^{(5)}R_{\rho\sigma}n^\sigma q^{~\rho}_\mu \,,
\label{Codacci}
\end{equation}
where the extrinsic curvature  of $M$ is denoted by 
$K_{\mu\nu}= q_\mu^{~\alpha} q_\nu^{~\beta} \nabla_\alpha n_\beta$, 
$K=K^\mu_\mu$ is its trace, and $D_\mu$ is 
the covariant differentiation with respect to $q_{\mu\nu}$.
Contracting the Gauss equation~(\ref{Gauss}) on 
$\alpha$ and $\gamma$, we find
\begin{equation}
{}^{(4)}R_{\mu\nu}
= {}^{(5)}R_{\rho\sigma} q_\mu^{~\rho}q_\nu^{~\sigma} -  
{}^{(5)}R^\alpha_{~\beta\gamma\delta}n_\alpha q_\mu^{~\beta} n^\gamma 
q_\nu^{~\delta}  + KK_{\mu\nu} 
-K^{~\alpha}_{\mu}K_{\nu\alpha}\,. 
\label{Ricci}
\end{equation}
This readily gives
\begin{equation}
{}^{(4)}G_{\mu\nu}
= \left[{}^{(5)}R_{\rho\sigma} -
{1 \over 2} g_{\rho\sigma}{}^{(5)}R\right] q_\mu^{~\rho} q_\nu^{~\sigma} 
+{}^{(5)}R_{\rho\sigma}n^\rho n^\sigma q_{\mu\nu}  + KK_{\mu\nu} 
-K^{~\rho}_{\mu}K_{\nu\rho} -{1 \over 2}q_{\mu\nu}  (K^2 
-K^{\alpha\beta}K_{\alpha\beta}) - \tilde{E}_{\mu\nu}\,, 
\label{4dEinstein-1}
\end{equation}
where 
\begin{equation}
\tilde{E}_{\mu\nu} 
\equiv {}^{(5)}R^\alpha_{~\beta\rho\sigma}n_\alpha n^\rho 
q_\mu^{~\beta} q_\nu^{~\sigma} \,.
\end{equation}
Using the 5-dimensional Einstein equations,
\begin{equation}
{}^{(5)}R_{\alpha\beta}
-\frac{1}{2}g_{\alpha\beta}{}^{(5)}R
=\kappa_5^2\, T_{\alpha\beta}\, ,
\label{5dEinstein}
\end{equation}
where  $T_{\mu\nu}$ is the 5-dimensional energy-momentum tensor, 
together with the decomposition of the Riemann tensor into the Weyl
curvature, the Ricci tensor and the scalar curvature; 
\begin{eqnarray}
{}^{(5)}R_{\mu\alpha\nu\beta}
=\frac{2}{3}( g_{\mu [\nu}{}^{(5)}R_{\beta]\alpha}
-g_{\alpha [ \nu}{}^{(5)}R_{\beta] \mu})
-\frac{1}{6}g_{\mu [\nu}g_{\beta ]\alpha}{}^{(5)}R
+{}^{(5)}C_{\mu\alpha\nu\beta}, 
\end{eqnarray}
we obtain the 4-dimensional equations as
\begin{equation}
{}^{(4)}G_{\mu\nu}
= {2 \kappa_5^2 \over 3}\left(T_{\rho\sigma}
q^{~\rho}_{\mu}  q^{~\sigma}_{\nu}  
+\left(T_{\rho\sigma}n^\rho n^\sigma-{1 \over 4}T^\rho_{~\rho}\right)
 q_{\mu\nu} \right) 
+ KK_{\mu\nu} 
-K^{~\sigma}_{\mu}K_{\nu\sigma} -{1 \over 2}q_{\mu\nu}
  \left(K^2-K^{\alpha\beta}K_{\alpha\beta}\right) - E_{\mu\nu}, 
\label{4dEinstein}
\end{equation}
where
\begin{equation}
E_{\mu\nu} \equiv {}^{(5)}C^\alpha_{~\beta\rho\sigma}n_\alpha n^\rho 
q_\mu^{~\beta} q_\nu^{~\sigma} .
\label{Edef}
\end{equation}
Note that $E_{\mu\nu}$ is traceless.
{}From the Codacci equation~(\ref{Codacci}) 
and the 5-dimensional Einstein equations~(\ref{5dEinstein}), 
we find 
\begin{equation}
D_\nu K^{~\nu}_\mu - D_\mu K = \kappa_5^2\, T_{\rho\sigma}
n^\sigma q_\mu^{~\rho} .
\label{momentum}
\end{equation}

So far we have not assumed any particular symmetry nor particular
form of the energy momentum tensor.
{}From now on, we take a brane world scenario. 
For convenience, we choose a coordinate $\chi$ such that
the hypersurface $\chi=0$ coincides with the brane world and
$n_\mu dx^\mu=d\chi$, which implies
\begin{eqnarray}
a^\mu=n^\nu\nabla_\nu n^\mu=0\,.
\end{eqnarray}
This is a condition on the coordinate in the direction of the extra
dimension. We assume this choice is possible at least in the
neighbourhood of the brane, $(M,q_{\mu\nu})$. In more explicit terms,
we assume the 5-dimensional metric to have the form,
\begin{equation}
ds^2=d\chi^2+q_{\mu\nu}dx^\mu dx^\nu\,.
\end{equation}

Bearing brane world spirit in mind, we 
assume that the 5-dimensional energy-momentum tensor has the form
%
\begin{eqnarray}
T_{\mu\nu}=-\Lambda g_{\mu\nu}+S_{\mu\nu}\delta (\chi), \label{eq:bulk}
\end{eqnarray}
%
where
%
\begin{eqnarray}
S_{\mu\nu}=-\lambda q_{\mu\nu}+\tau_{\mu\nu}\,, \label{eq:matter}
\end{eqnarray}
%
with $\tau_{\mu\nu}n^\nu=0$. 
$\Lambda$ is the 
cosmological constant of the bulk spacetime. 
$\lambda$ and $\tau_{\mu\nu}$ are the
vacuum energy and the energy-momentum tensor, respectively,
in the brane world. Note that $\lambda$ is the tension
of the brane in 5 dimensions.
Properly speaking $S_{\mu\nu}$ should be evaluated by the
variational principle of the 4-dimensional Lagrangian for matter
fields because the normal matter except for gravity is assumed to be 
living only in the $\chi=0$ brane.  It should be noted that
the decomposition of $S_{\mu\nu}$ into $\lambda q_{\mu\nu}$ and
$\tau_{\mu\nu}$ can be ambiguous, particularly in cosmological
contexts. 

The singular behaviour in the energy-momentum tensor leads us 
to the so-called Israel's junction condition\cite{israel}, 
%
\begin{eqnarray}
\left[q_{\mu\nu}\right]
&=&0\,,
\nonumber\\
\left[K_{\mu\nu}\right]&=&
-\kappa_5^2 \Bigl(S_{\mu\nu}-\frac{1}{3}q_{\mu\nu}S\Bigr),
\end{eqnarray}
%
where $[X]:=\lim_{\chi \to +0}X - \lim_{\chi \to -0}X=X^+-X^-$. 

Now we impose the $Z_2$-symmetry on this spacetime, with the brane as
the fixed point. Interestingly the symmetry uniquely determines the
extrinsic curvature of the brane in terms of the energy momentum tensor,
%
\begin{eqnarray}
K_{\mu\nu}^+=-K^-_{\mu\nu}=-\frac{1}{2}\kappa_5^2 
\Bigl(S_{\mu\nu}-\frac{1}{3}q_{\mu\nu}S\Bigr)\,. \label{eq:zsym}
\end{eqnarray}
%
Hereafter we focus our attention on quantities evaluated on the brane.
Because of the $Z_2$-symmetry, we may evaluate quantities either on the
$+$ or $-$ side of the brane. Hence we omit the indices $\pm$ below for
brevity.

Substituting Eq.~(\ref{eq:zsym}) into Eq.~(\ref{4dEinstein}),
we obtain the gravitational equations on the 3-brane in the form,
%
\begin{eqnarray}
{}^{(4)}G_{\mu\nu}=-\Lambda_4q_{\mu\nu}
+ 8 \pi G_N\tau_{\mu\nu}+\kappa_5^4\,\pi_{\mu\nu}
-E_{\mu\nu}\,, \label{eq:effective}
\end{eqnarray}
%
where
\begin{eqnarray}
\Lambda_4&=&\frac{1}{2}\kappa_5^2
\left(\Lambda +\frac{1}{6}\kappa_5^2\,\lambda^2\right)\,,
\label{Lamda4}\\
G_N&=&{\kappa_5^4\,\lambda\over48 \pi}\,,
\label{GNdef}\\
\pi_{\mu\nu}&=&
-\frac{1}{4} \tau_{\mu\alpha}\tau_\nu^{~\alpha}
+\frac{1}{12}\tau\tau_{\mu\nu}
+\frac{1}{8}q_{\mu\nu}\tau_{\alpha\beta}\tau^{\alpha\beta}-\frac{1}{24}
q_{\mu\nu}\tau^2\,,
\label{pidef}
\end{eqnarray}
and $E_{\mu\nu}$ is the part of the 5-dimensional Weyl tensor defined
in Eq.~({\ref{Edef}). It should be noted that $E_{\mu\nu}$ in the above
is the limiting value at $\chi=+0$ or $-0$ but not the value exactly
on the brane. This is our main result. It resembles the conventional
Einstein equations in 4 dimensions. 
In fact, the Einstein equations can be recovered by taking the 
limit $\kappa_5\to0$ while keeping $G_N$ finite.
Nevertheless there are some important differences.
As can be easily seen, the existence of Newton's
gravitational constant $G_N$ strongly relies on the presence of the
vacuum energy $\lambda$. In other words, it becomes impossible to define
Newton's gravitational constant during an era when the distinction
between the vacuum energy and the normal matter energy is ambiguous.
Furthermore, we would have the wrong sign of $G_N$ if 
$\lambda<0$\cite{Csaba}.
The $\pi_{\mu\nu}$ term, which is quadratic in $\tau_{\mu\nu}$ 
could play a very important role, especially in the early universe
when the matter energy scale is high\cite{BDL,Csaba,Cline}.

In addition to these features that have been pointed out previously,
Eq.~(\ref{eq:effective}) contains a new term, $E_{\mu\nu}$.
It is a part of the 5-dimensional Weyl tensor and carries information
of the gravitational field outside the brane.
It is non-vanishing if the bulk spacetime is not purely anti-de Sitter.
At the same time, it is not freely specifiable
but is constrained by the motion of the matter on the brane.
Let us show this feature now.
Together with Eq.~(\ref{eq:zsym}), 
Eq.~(\ref{momentum}) implies the conservation law for the
matter, 
%
\begin{eqnarray}
D_\nu K^{~\nu}_\mu - D_\mu K \propto
D_\nu \tau^{~\nu}_{\mu}=0.
\end{eqnarray}
%
Therefore the contracted Bianchi identities 
$D^\mu {}^{(4)}G_{\mu\nu}=0$
imply the relation between $E_{\mu\nu}$ and $\tau_{\mu\nu}$ as
%
\begin{eqnarray}
D^\mu E_{\mu\nu} & = & K^{\alpha\beta}(D_\nu K_{\alpha\beta}-D_\beta 
K_{\nu \alpha} ) \nonumber \\
& = & \frac{1}{4}\kappa_5^4 \Bigl[ \tau^{\alpha\beta}
(D_\nu \tau_{\alpha\beta}-D_\beta \tau_{\nu\alpha})+\frac{1}{3}
(\tau_{\mu\nu}-q_{\mu\nu}\tau )D^\mu \tau \Bigr]. 
 \label{eq:cond}
\end{eqnarray}
%
Thus $E_{\mu\nu}$ is not freely specifiable but its divergence
is constrained by the matter term. If one further decomposes
$E_{\mu\nu}$ into the transverse-traceless part, $E_{\mu\nu}^{TT}$,
and the longitudinal part, $E_{\mu\nu}^{L}$, 
the latter is determined completely by the matter. 
Hence if the $E_{\mu\nu}^{TT}$ part is absent, the equations will
be closed solely with quantities that reside in the brane. However,
as usually the case in the conventional gravity, the $E_{\mu\nu}^{TT}$
part corresponds to gravitational waves or gravitons in 5 dimensions,
and they will be inevitably excited by matter motions and 
their excitations affect matter motions in return.
This implies the effective gravitational equations on the brane are
not closed but one must solve the gravitational field in the bulk at
the same time in general. Since the derivation of equations that
govern the evolution of $E_{\mu\nu}^{TT}$ is technically
complicated, we defer it to Appendix A.

Let us now estimate the effect of each term on the right-hand side of
Eq.~(\ref{eq:effective}). We set $\kappa_5^{-2}=M_G^3$ and
$\lambda=M_\lambda^4$, and assume $\Lambda=O(\kappa_5^2\,\lambda^2)$.
It should be noted that these do not have to be planck scale quantities.
One can scale them as $M_G\to f^2M_G$ and 
$M_\lambda\to f^3M_\lambda$, where $f$ is an arbitrary constant, while
keeping the gravitational constant $G_N$ unchaged. 
Nevertheless, here we assume $M_G$ and $M_\lambda$ to be sufficiently
large compared to the characteristic energy scale of the matter which we
denote by $M$.

The first term on the right-hand side of
Eq.~(\ref{eq:effective}) is the net cosmological constant
in 4 dimensions. It is assumed that $\Lambda<0$.
Hence $\Lambda_4$ may take arbitrary value as one may wish by
appropriately specifying the values of $\Lambda$ and $\lambda$.
The second term is the contribution from normal matters which should
satisfy the local energy condition (assuming the decomposition
of $S_{\mu\nu}$ into $\lambda$ and $\tau_{\mu\nu}$ is well-defined).
The $\pi_{\mu\nu}$ term which is quadratic in $\tau_{\mu\nu}$ 
is expected to be negligible in the low energy limit. 
In fact, the ratio of these terms to the third term is 
approximately given by 
%
\begin{eqnarray}
{\kappa_5^4\,|\pi_{\mu\nu}|\over G_N|\tau_{\mu\nu}|}
\sim\frac{\kappa_5^4\,|\tau_{\mu\alpha} \tau_\nu^{~\alpha}+\cdots|}
{G_N |\tau_{\mu\nu}|} \sim \frac{M^4}{M_{\lambda}^4}\,.
\end{eqnarray}
%

We now turn to the Weyl tensor part. First let us consider
the longitudinal part $E_{\mu\nu}^{L}$. Since it is
determined by $\tau_{\mu\nu}$ through Eq.~(\ref{eq:cond}),
we have
%
\begin{eqnarray}
\frac{|E_{\mu\nu}^{L}|}{G_N|\tau_{\mu\nu}|} \sim 
\frac{\kappa_5^4\,|\tau_{\mu\alpha} \tau_\nu^{~\alpha}+\cdots|}
{G_N |\tau_{\mu\nu}|} \sim 
\frac{M^4}{M_{\lambda}^4}\,.  
\end{eqnarray}
%
This is the same order of magnitude as the $\pi_{\mu\nu}$ term.
Second, we consider the $E_{\mu\nu}^{TT}$ part.
We focus on the effect due to matter excitations
on the brane.  Here we borrow the discussion of \cite{RS2} to evaluate
the leading order of magnitude of its effect. 
The gravitational potential between two bodies on
the brane is modified via exchange of gravitons living in 5
dimensions as\cite{RS2} 
%
\begin{eqnarray}
V(r) \sim \frac{G_Nm_1 m_2}{r}\Bigl( 1+\frac{1}{r^2 k^2}\Bigr), 
\end{eqnarray}
%
where $r$ is the distance between the two bodies and
$k=\kappa_5^2\,\lambda/6$. 
Since this effect must be contained in Eq.~(\ref{eq:effective}),
it should appear in the $E_{\mu\nu}$ term.
Therefore as a conservative estimate, we obtain
%
\begin{eqnarray}
\frac{|E_{\mu\nu}|}{G_N|\tau_{\mu\nu}|} \sim 
\frac{M^2}{k^2}\sim{M_G^6M^2\over M_\lambda^8}\,.
\end{eqnarray}
%
Thus $E_{\mu\nu}$ is also negligible in the low energy world. It is,
however, worth noting that this term is larger than the terms quadratic
in $\tau_{\mu\nu}$. The deviation from the ordinary Einstein equations
in 4-dimensions first appears from gravitational excitations in
the bulk spacetime. {}From the above estimations we conclude that the
effective gravitational equation~(\ref{eq:effective}) on the brane 
reduce to the 4-dimensional conventional Einstein gravity, 
${}^{(4)}G_{\mu\nu} \simeq -{\Lambda_4}q_{\mu\nu}+8 \pi G_N
\tau_{\mu\nu}$, in the low energy limit. The presence of a well-defined
cosmological constant
$\lambda$ is obviously essential here.\footnote{As we can
see from the first term in Eq.~(\ref{4dEinstein}), 
the reduction to the normal Einstein gravity is also possible
with the introduction of non-trivial bulk 
energy-momentum tensor\cite{Olive}. }

Finally, we note an outcome of the constraint~(\ref{eq:cond}).
We consider the case when the bulk
spacetime is pure anti-de Sitter with $E_{\mu\nu}=0$ and
investigate the condition on the matter on the brane.
For simplicity, we assume the perfect fluid form for the
energy momentum tensor:
%
\begin{eqnarray}
\tau^{\mu\nu}=\rho\, t^\mu t^\nu+Ph^{\mu\nu}\,,
\end{eqnarray}
where $h^{\mu\nu}=q^{\mu\nu}+t^\mu t^\nu$. 
The quadratic term $\pi^{\mu\nu}$ in the 4-dimensional
effective gravitational equations~(\ref{eq:effective}) then becomes
%
\begin{eqnarray}
\pi^{\mu\nu} = \frac{1}{12}\rho\left(\rho\, t^\mu t^\nu
+(\rho +2P)h^{\mu\nu}\right)\,.
\end{eqnarray}
%
The normal conservation law $D_\nu \tau^{\mu\nu}=0$ implies 
\begin{eqnarray}
t^\mu D_\mu\rho +(\rho +P)D_\mu t^\mu=0~~~{\rm and}~~~
(\rho+P)t^\nu D_\nu t^\mu +h^{\mu\nu}D_\nu P=0\,. 
\label{eq:bian}
\end{eqnarray}
If $E_{\mu\nu}=0$, the 4-dimensional Bianchi identities
imply $D_\nu \pi^{\mu\nu}=0$, which gives
%
\begin{eqnarray}
D_\nu \pi^{\mu\nu}
=\frac{1}{6}(\rho +P)h^{\mu\nu}D_\nu \rho=0\,.
\end{eqnarray}
%
This means $\partial_i\rho=0$. Hence an inhomogeneous 
perfect fluid is rejected.

We briefly summarize the present work.  
We first derived the effective 4-dimensional gravitational equations in 
5 dimensions, Eq.~(\ref{4dEinstein}), without any particular
assumptions specific to the brane world scenario.
Then based on the brane world scenario, we introduced the $Z_2$ symmetry
and assumed that the matter lives only 
on the brane, and derived the 4-dimensional effective gravitational
equations on the brane, Eq.~(\ref{eq:effective}). The 
equation tells us that a normal gravitational theory can be obtained
on the brane only if the tension is positive, while
an RS1-type theory\cite{RS1} in which the brane has negative tension 
is rejected from the physical point of view (see also \cite{Csaba} 
for Friedmann cases). 
In the case of the brane with positive tension, the Einstein gravity is
recovered in the low energy limit. 
Placing the brane in the 5-dimensional exact 
anti-de Sitter spacetime imposes a strong condition on the matter 
in 4-dimensions. In particular, if the matter energy-momentum tensor
has the perfect fluid form, only spatially
homogeneous universes are allowed.  Conversely, this means that the
deviation of the bulk spacetime from the exact anti-de Sitter spacetime
is essential to describe our real world with matter fields.

\section*{Acknowledgements}
TS would like to thank Gary W. Gibbons and DAMTP 
relativity group for their hospitality at Cambridge. 
Substantial part of this work was done while 
KM and MS were participating the program, ``Structure Formation in the
Universe'', at the Newton Institute, University of Cambridge.
We are grateful to the Newton Institute for their hospitality. 
TS's work is supported by JSPS Postdoctal Fellowship for 
Research Abroad. KM's work is supported in part by Monbusho
Grant-in Aid for Specially Promoted Research No.~08102010.
MS's work is supported in part by Monbusho Grant-in-Aid
for Scientific Research No.~09640355.

\appendix

\section{}
We derive the evolution equation of ${E}_{\mu\nu}$ 
to make our system of equations closed. 
First, we write down the Weyl tensor 
formulas. The $n$-dimensional Riemann tensor is written in terms of 
the Weyl and Ricci tensors as
%
\begin{eqnarray}
{}^{(n)}R_{\alpha\beta\mu\nu}
={}^{(n)}C_{\alpha\beta\mu\nu}
+{2\over n-2}\left({}^{(n)}R_{\alpha[\mu}g_{\nu]\beta}
-{}^{(n)}R_{\beta[\mu}g_{\nu]\alpha}\right)
-{2\over(n-1)(n-2)}{}^{(n)}R\,g_{\alpha[\mu}g_{\nu]\beta}.
\label{Riemann}
\end{eqnarray}
%
We decompose the Weyl tensor into the `electric' and `magnetic' parts:
%
\begin{eqnarray}
E_{\mu\nu}\equiv{}^{(n)}C_{\mu\alpha\nu\beta}n^\alpha n^\beta\,,
\end{eqnarray}
%
and
%
\begin{eqnarray}
B_{\mu\nu\alpha}\equiv q_\mu^\rho q_\nu^\sigma
{}^{(n)}C_{\rho\sigma\alpha\beta}n^\beta\,.
\end{eqnarray}
%
$B_{\mu\nu\alpha}$ and $E_{\mu\nu}$ have the symmetry,
%
\begin{eqnarray}
&&B_{\alpha\beta\mu}=-B_{\beta\alpha\mu}\,,
\quad B_{[\alpha\beta\mu]}=0\,,
\quad B^{\alpha}{}_{\beta\alpha}=0\,.
\nonumber\\
&&E_{\alpha\beta}=E_{\beta\alpha}\,,
\quad E^{\alpha}{}_\alpha=0\,.
\end{eqnarray}
%
The algebraic degrees of freedom are 
%
\begin{eqnarray}
&&{}^{(n)}R_{\alpha\beta\mu\nu}~\cdots~{n^2(n^2-1)\over12}\,,
\quad
{}^{(n)}C_{\alpha\beta\mu\nu}~\cdots~{(n-3)n(n+1)(n+2)\over12}\,,
\nonumber\\
&&{}^{(n)}R_{\mu\nu}~\cdots~{n(n+1)\over2}\,,
\quad
{}^{(n-1)}C_{\alpha\beta\mu\nu}~\cdots~{(n-4)(n-1)n(n+1)\over12}\,,
\nonumber\\
&&B_{\alpha\beta\mu}~\cdots~{(n-3)(n-1)(n+1)\over3}\,,
\quad
E_{\alpha\beta}~\cdots~{(n-2)(n+1)\over2}\,.
\end{eqnarray}
%
The $n$-dimensional Weyl tensor can be written in terms of 
${}^{(n-1)}C_{\alpha\beta\mu\nu}$, $E_{\alpha\beta}$, 
$B_{\mu\nu\alpha}$ and the extrinsic curvature $K_{\mu\nu}$, 
%
\begin{eqnarray}
{}^{(n)}C_{\alpha\beta\mu\nu}
=&&{}^{(n-1)}C_{\alpha\beta\mu\nu}
+2B_{\alpha\beta[\mu}n_{\nu]}+2B_{\mu\nu[\alpha}n_{\beta]}
\nonumber\\
&&
+\left(2E_{\alpha[\mu}n_{\nu]}n_\beta
                     -2E_{\beta[\mu}n_{\nu]}n_\alpha\right)
-{1\over n-3}\left(2E_{\alpha[\mu}q_{\nu]\beta}
                     -2E_{\beta[\mu}q_{\nu]\alpha}\right)
\nonumber\\
&&
-f_{\alpha\beta\mu\nu}
+{2\over n-3}\left(q_{\alpha[\mu}f_{\nu]\beta}
-q_{\beta[\mu}f_{\nu]\alpha}\right)
-{2\over(n-2)(n-3)}f^\sigma{}_\sigma q_{\alpha[\mu}q_{\nu]\beta}\,,
\label{weyldecomp}
\end{eqnarray}
%
where $q_{\mu\nu}=g_{\mu\nu}-n_\mu n_\nu$ and
%
\begin{eqnarray}
f_{\alpha\beta\mu\nu}\equiv
&&K_{\alpha\mu}K_{\beta\nu}-K_{\alpha\nu}K_{\beta\mu}\,,
\nonumber\\
f_{\mu\nu}\equiv&&f_{\mu}{}^{\sigma}{}_{\nu\sigma}
=f{}^{\sigma}{}_{\mu\sigma\nu}
=KK_{\mu\nu}-K_{\mu\sigma}K^{\sigma}{}_{\nu}=f_{\nu\mu}\,,
\nonumber\\
f^\mu{}_\mu=&&f^{\mu\nu}{}_{\mu\nu}=K^2-K^{\mu\nu}K_{\mu\nu}\,.
\label{KKdef}
\end{eqnarray}
%

{}From now on we set $n=5$ and derive the evolution equations
of $E_{\mu\nu}$ from the 5-dimensional Bianchi identities.
We assume $a^\mu=n^\alpha\nabla_\alpha n^\mu=0$. 
For convenience, we define ${\tilde E}_{\mu\nu}$ and ${\tilde
B}_{\mu\nu\alpha}$ from the Riemann tensor,
%
\begin{eqnarray}
{\tilde E}_{\mu\nu}&\equiv&{}^{(5)}R_{\mu \alpha\nu\beta}n^\alpha n^\beta
=-\mbox \pounds_n K_{\mu\nu}+K_{\mu\alpha}K_\nu^\alpha\,,
\label{tildeE}\\
{\tilde B}_{\mu\nu\alpha}&\equiv&q_\mu^\beta q_\nu^\sigma 
{}^{(5)}R_{\beta\sigma \alpha \rho} n^\rho=2D_{[\mu}K_{\nu]\alpha}\,.
\label{tildeB}
\end{eqnarray}
%
These are related to $E_{\mu\nu}$ and $B_{\mu\nu\alpha}$ as
%
\begin{eqnarray}
E_{\mu\nu} & = &
 {\tilde E}_{\mu\nu}-\frac{1}{3}q_{\mu\nu}{}^{(5)}R_{\alpha\beta}
n^\alpha n^\beta
-\frac{1}{3}q_\mu^\alpha q_\nu^\beta {}^{(5)}R_{\alpha\beta}
+\frac{1}{12}q_{\mu\nu}{}^{(5)}R 
\nonumber \\
& = & -\frac{1}{3}\Bigl({}^{(4)}R_{\mu\nu}
-\frac{1}{4}q_{\mu\nu}{}^{(4)}R \Bigr)
-\frac{2}{3} \mbox \pounds_n \Bigl(K_{\mu\nu}-\frac{1}{4}q_{\mu\nu}K \Bigr) 
+\frac{1}{3}K_{\mu\alpha}K^\alpha_\nu+\frac{1}{4}q_{\mu\nu}
\Bigl(K_{\alpha\beta}K^{\alpha\beta}-\frac{1}{3}K^2 \Bigr)\,,
\label{EE}\\
B_{\mu\nu\alpha} & = & {\tilde B}_{\mu\nu\alpha}+\frac{2}{3}
(D_\beta K^\beta_{[\mu}-D_{[\mu}K)q_{\nu]}^\alpha 
\nonumber \\
& = & 2D_{[\mu}K_{\nu]\alpha}+\frac{2}{3}
(D_\beta K^\beta_{[\mu}-D_{[\mu}K)q_{\nu]}^\alpha \,.
\label{eq:bdeco}
\end{eqnarray}
%

The 5-dimensional Bianchi identities are
%
\begin{eqnarray}
\nabla_{[\mu} {}^{(5)}R_{\nu \alpha]\beta \sigma}=0\,,
\end{eqnarray}
%
from which we obtain the following four sets of identities:
%
\begin{eqnarray}
D_{[\mu}{\tilde B}_{\nu \alpha]}^{~~~\beta}+K^\sigma_{[\mu}
{}^{(4)}R_{\nu \alpha] \sigma}^{~~~~~\beta}
=0\,,
\label{eq:bianchi1}
\end{eqnarray}
%
%
\begin{eqnarray}
\mbox \pounds_n {\tilde B}_{\mu\nu\alpha}
+2D_{[\mu}{\tilde E}_{\nu]\alpha}-K_\alpha^\sigma 
{\tilde B}_{\mu\nu\sigma}
+2{\tilde B}_{\alpha \sigma [\mu }K_{\nu]}^\sigma
=0\,,
\label{eq:bianchi2}
\end{eqnarray}
%
%
\begin{eqnarray}
\mbox \pounds_n {}^{(4)}R_{\mu \nu \alpha \beta}
+2{}^{(4)}R_{\mu\nu\sigma [\alpha}K^\sigma_{\beta]} 
+2D_{[\mu}{\tilde B}_{|\alpha\beta|\nu]}
=0\,,
\label{eq:bianchi3}
\end{eqnarray}
%
%
\begin{eqnarray}
D_{[\mu}{}^{(4)}R_{\nu \alpha ] \beta \sigma}=0\,.
\label{eq:bianchi4}
\end{eqnarray}
%

{}From Eq.~(\ref{eq:bdeco}) and the Israel's junction condition, we
obtain
%
\begin{eqnarray}
[{\tilde B}_{\mu\nu \alpha}]=2D_{[\mu}[K_{\nu]\alpha}]=
-2\kappa_5^2D_{[\mu}\Bigl( \tau_{\nu] \alpha}
-\frac{1}{3}q_{\nu] \alpha}\tau \Bigr).
\end{eqnarray}
%
Thus the $Z_2$-symmetry uniquely determines the value of
$B_{\mu\nu\alpha}$ on the brane as
%
\begin{eqnarray}
{\tilde B}_{\mu\nu \alpha}^+
&=&-{\tilde B}_{\mu\nu \alpha}^-
\nonumber\\
&=&-\kappa_5^2D_{[\mu}\Bigl( \tau_{\nu] \alpha}
-\frac{1}{3}q_{\nu] \alpha}\tau \Bigr),
\nonumber\\
{B}^+_{\mu\nu \alpha} & = & 2D_{[\mu}K^+_{\nu]\alpha}
+\frac{2}{3}
\Bigl( D_\beta K^{+\beta}_{[\mu}-D_{[\mu}K^+ \Bigr)q_{\nu]\alpha}
\nonumber \\
& =& {\tilde B}_{\mu\nu \alpha}^+\,.
\end{eqnarray}
%
These equations give the boundary conditions on the brane when
one solves the evolution of $E_{\mu\nu}$ in 5 dimensions.

The equations that govern the evolution of $E_{\mu\nu}$
in the bulk (i.e., in the spacetime region away from the brane) are
obtained as follows. Using the 5-dimensional Einstein 
equations~(\ref{5dEinstein}), Eq.~(\ref{eq:bianchi2}) yields 
%
\begin{eqnarray}
\mbox \pounds_n {B}_{\mu\nu\alpha}=-
2D_{[\mu}{E}_{\nu]\alpha}+K_\alpha^\sigma 
{B}_{\mu\nu\sigma}
-2{B}_{\alpha \sigma [\mu }K_{\nu]}^\sigma\,,
\label{eq:BBB}
\end{eqnarray}
%
in the bulk. 
Also, using Eq.~(\ref{5dEinstein}) and (\ref{4dEinstein}), 
Eq.~(\ref{eq:bianchi3}) gives
%
\begin{eqnarray}
\mbox \pounds_n E_{\alpha\beta} = &&
D^\mu B_{\mu(\alpha\beta)}
+ \frac{1}{6}\kappa_5^2 
\Lambda\left(K_{\alpha\beta}-q_{\alpha\beta}K\right)
+K^{\mu\nu}{}^{(4)}R_{\mu\alpha\nu\beta}
 \nonumber \\
&& +3K^\mu_{(\alpha}E_{\beta)\mu}-KE_{\alpha\beta}
+\left(K_{\alpha\mu}K_{\beta\nu}
-K_{\alpha\beta}K_{\mu\nu}\right)K^{\mu\nu}\,,
\label{eq:EEE}
\end{eqnarray}
in the bulk. 
Together with the 4-dimensional Einstein equations~(\ref{4dEinstein})
 in the bulk, Eqs.~(\ref{eq:BBB}) and (\ref{eq:EEE})
form a closed system of equations. In particular, one may easily
recognize the wave-like character of the transverse part of
$E_{\mu\nu}$, which propagates as gravitons in 5 dimensions.

\end{document}